\documentclass{article}

\usepackage{physics}
\usepackage{amsmath}
\usepackage{amsfonts}
\usepackage[icelandic,english]{babel}
\usepackage[T1]{fontenc}
\usepackage[utf8]{inputenc}
\usepackage{jheppub}
\usepackage{appendix}
\usepackage{color}
\newcommand{\pdd}{\partial}
\newcommand{\covd}{\nabla}

\newcommand{\mc}{\mathcal}

\newcommand{\bbR}{\mathbb{R}}
\newcommand{\bbZ}{\mathbb{Z}}

\newcommand{\rp}{r_{+}}
\newcommand{\rmm}{r_{-}}
\newcommand{\rhp}{\rho_{+}}
\newcommand{\rhm}{\rho_{-}}

\newcommand{\rt}{\tilde{r}}
\newcommand{\rpt}{\tilde{r}_{+}}
\newcommand{\rmt}{\tilde{r}_{-}}
\newcommand{\taut}{\tilde{\tau}}
\newcommand{\thetat}{\tilde{\theta}}
\newcommand{\trhp}{\tilde{\rho}_+}
\newcommand{\trhm}{\tilde{\rho}_-}

\title{A Selberg zeta function for warped AdS$_3$ black holes}

\author{Victoria L. Martin,}
\author{Rahul Poddar}
\author{and Agla Þórarinsdóttir}

\affiliation{Science Institute, University of Iceland,\\Dunhaga 3, 107 Reykjavík, Iceland}

\emailAdd{vlmartin@hi.is}
\emailAdd{rap19@hi.is}
\emailAdd{agt15@hi.is}

\abstract{ 
 The Selberg zeta function and trace formula are powerful tools used to calculate kinetic operator spectra and quasinormal modes on hyperbolic quotient spacetimes. In this article, we extend this formalism to non-hyperbolic quotients by constructing a Selberg zeta function for warped AdS$_3$ black holes. We also consider the so-called self-dual solutions, which are of interest in connection to near-horizon extremal Kerr. We establish a map between the zeta function zeroes and the quasinormal modes on warped AdS$_3$ black hole backgrounds. In the process, we use a method involving conformal coordinates and the symmetry structure of the scalar Laplacian to construct a warped version of the hyperbolic half-space metric, which to our knowledge is new and may have interesting applications of its own, which we describe. We end by discussing several future directions for this work, such as computing 1-loop determinants (which govern quantum corrections) on the  quotient spacetimes we consider, as well as adapting the formalism presented here to more generic orbifolds.
}

\begin{document}
\maketitle

\section{Introduction}

%The Selberg zeta function and trace formula are now well-established tools in physics to learn about dynamics and quantum corrections on hyperbolic quotient

By studying only the quotient structure of certain spacetimes, it is possible to learn much about dynamics and quantum corrections on those backgrounds. A principal tool to accomplish this is the Selberg zeta function, a cousin of the Riemann zeta function in which prime numbers are replaced by prime geodesics on a hyperbolic spacetime $\mathbb{H}^n/\Gamma$, where $\Gamma$ is a discrete subgroup of SL(2,$\mathbb{R}$) \cite{1570009749218165120,perry2003poisson,perry2003selberg}. For example, for $\mathbb{H}^2/\Gamma$ the Selberg zeta function is of the form
\begin{equation}\label{selberggeneral}
    Z_\Gamma(s)=\prod_p\prod_{n=0}^{\infty}\left(1-N(p)^{-s-n}\right),
\end{equation}
where the product $p$ is over the conjugacy classes of prime geodesics, and $N(p)$ is a function of geodesic length \cite{hejhal548selberg}. The Selberg zeta function and trace formula are of significant physical interest as a tool to compute spectra of kinetic operators (and thus quantum corrections) on hyperbolic quotient manifolds. A brief list of applications includes quantum chaos \cite{Balazs:1986uj,marklof2004selberg}, quantum JT gravity \cite{Garcia-Garcia:2019zds}, torsion and topological invariants \cite{Bytsenko:1997sr,Bytsenko:1998uy,Bytsenko:2001xs}, heat kernels and regularized one-loop determinants \cite{Bytsenko:1994bc,giombi2008one,Bytsenko:2008wj,David:2009xg,Keeler:2018lza}, quantum corrections to black hole entropy \cite{Bytsenko:1997ru,Bytsenko:2013psa}, quasinormal modes \cite{Keeler:2018lza,Aros:2009pg}, and even band theory \cite{attar2022selberg}.

So far, the Selberg zeta function formalism has been limited to hyperbolic quotient spacetimes. A notable example is the Euclidean BTZ black hole \cite{banados1992black}, which is a quotient of hyperbolic 3-space $\mathbb{H}^3$ by the discrete subgroup $\Gamma\sim\mathbb{Z}$ \cite{Banados:1992gq}. Perry and Williams \cite{perry2003selberg} constructed a Selberg zeta function for the BTZ black hole from this quotient structure, and proved a corresponding trace formula. In that work, the action of the quotienting group $\Gamma$ is studied through a set of ``conformal coordinates'' that transform the BTZ metric in Boyer-Lindquist-like coordinates to the Poincar\'e patch. It was discussed in \cite{perry2003selberg,Aros:2009pg} that the zeroes of the Selberg zeta function $\left\{s_*|Z_\Gamma(s_*)=0\right\}$ can be mapped to the BTZ quasinormal modes. This was worked out explicitly in \cite{Keeler:2018lza}, where it was shown that tuning the Selberg zeroes $s_*$ to the conformal dimension of the field in question was equivalent to equating the field's quasinormal modes to the thermal (Matsubara) frequencies:
\begin{equation}
    s_*=\Delta ~~\leftrightarrow~~ \omega_{QN}=\omega_n.
\end{equation}
This analysis was extended to fields of general spin in \cite{Keeler:2019wsx}.

In this work, we construct a Selberg zeta function for warped AdS$_3$ black holes, extending the work of \cite{perry2003selberg} beyond hyperbolic quotients for the first time. Warped AdS$_3$ black holes are quotients of warped AdS$_3$ \cite{Anninos:2008fx}, with quotienting group $\Gamma\sim\mathbb{Z}$ realized by the discrete identification $\phi\rightarrow\phi+2\pi n$, with $n$ an integer. The isometry group is SL(2,$\mathbb{R})\times$U(1). Warped AdS$_3$ geometries arise in many different contexts in theoretical physics, such as topological massive gravity (see \cite{Anninos:2008fx} and references therein), lower-spin gravity \cite{Hofman:2014loa} (in the context of warped black holes, see \cite{Azeyanagi:2018har}), and in near-horizon extremal \cite{Guica:2008mu} and near-region non-extremal \cite{Castro:2010fd} Kerr. Thus it is very desirable to calculate quantum corrections on these backgrounds, and it is very likely that our formalism will help facilitate this calculation. This is addressed further in the discussion section.

Like \cite{perry2003selberg}, we build the Selberg zeta function using the quotient structure of warped AdS$_3$ black holes. However, unlike  \cite{perry2003selberg}, we do not have a straightforward way to analyze the group action under $\phi\rightarrow\phi+2\pi n$, because we do not have a warped version of the Poincar\'e patch as a target metric with which to construct conformal coordinates. We solve this problem in section \ref{section:SelberglikeWAdS}, where we use a trick from the hidden conformal symmetry program to exploit the symmetries of the scalar Laplacian to construct a set of conformal coordinates. Namely, we build conformal coordinates such that the SL(2,$\mathbb{R})\times$U(1) quadratic Casimir $\mathcal{H}^2 + \lambda H_0^2\propto\nabla^2$. Using the resulting conformal coordinate transformation, we are able to derive a warped version of the upper half-space metric that to our knowledge is new. This metric might have interesting applications of its own, such as the study of $T\overline{T}$ deformations of warped conformal field theories. We give more details about this is the in the discussion section.

We also consider the self-dual solutions reported in \cite{Anninos:2008fx} and further studied in \cite{Chen:2010qm,Li:2010sv,Li:2010zr}. These are the warped analogue of the self-dual solutions for AdS$_3$ presented in \cite{Coussaert:1994tu}, and as explained there these solutions are interesting to study in their own right. They are formed by a different quotient, $\tilde\theta\rightarrow\tilde\theta+2\pi\tilde\alpha n$, which we describe.

This article is organised as follows. In Section \ref{Sec2Review} we review previous results that will be helpful for the reader. In \ref{subsec:SelbergBTZ} we largely follow \cite{perry2003selberg} in demonstrating how to construct the Selberg zeta function for the case of the BTZ black hole, ending with mapping the Selberg zeroes to the BTZ quasinormal modes as done in \cite{Keeler:2018lza}. In \ref{subsec:2.2warpedbh} we review the quotient structure of warped AdS$_3$ black holes, largely following \cite{Anninos:2008fx}. In Section \ref{section:SelberglikeWAdS} we exploit the symmetries of the Klein-Gordon operator to build a set of conformal coordinates. These allow us to create both a warped Poincar\'e patch metric as well as the warped Selberg zeta function. In Section \ref{sec:wave} it is necessary to calculate several quantities on warped quotient backgrounds (namely the conformal weight $\Delta$, the thermal frequencies $\omega_n$ and the quasinormal mode frequencies $\omega_*$) before constructing the map between the Selberg zeroes and quasinormal mode frequencies in Section \ref{sec5:mappingzeroes}. In Section \ref{sec:discussion} we summarize our results and discuss several future directions.

\section{Review}\label{Sec2Review}
Here we review material that we will need throughout the rest of this work. We begin by reviewing how, in the case of the BTZ black hole, quotient structure can be used to construct a Selberg zeta function whose zeroes are mapped to the quasinormal modes on the BTZ background. We then review the warped AdS$_3$ black hole geometry and its quotient structure.  In Section \ref{section:SelberglikeWAdS} we put these two elements together and create a Selberg zeta function for warped AdS$_3$ black holes. 

\subsection{Constructing the Selberg zeta function for $\mathbb{H}^3/\Gamma$}\label{subsec:SelbergBTZ}
We begin with the  BTZ black hole metric in Boyer-Lindquist-like coordinates 

\begin{equation}\label{BTZmetric}
    ds^2=-\frac{(r^2-\rp^2)(r^2-\rmm^2)}{L^2r^2}dt^2 +\frac{L^2 r^2}{(r^2-\rp^2)(r^2-\rmm^2)}dr^2 + r^2\left(d\phi-\frac{r_+r_-}{Lr^2}dt\right)^2
\end{equation}
where $L$ is the AdS radius, and the outer and inner  horizons $\rp$ and $\rmm$ are related to the black hole's mass $M$ and angular momentum $J$ by $\rp^2+\rmm^2=L^2M$ and $\rp\rmm={J L}/{2}$. The Euclidean\footnote{The authors of \cite{perry2003selberg} work in Euclidean signature because the existence of well-defined Selberg zeta functions has been shown specifically for hyperbolic quotients. When we adapt this formalism to warped AdS$_3$ black holes in Section 3, we take the liberty of working in Lorentzian signature.} BTZ black hole is obtained by $t\rightarrow -i\tau$, $J\rightarrow -iJ_E$ and $r_-\rightarrow -i|r_-|$.

It is well-known \cite{Banados:1992gq} that the BTZ black hole metric can be mapped to the Poincar\'e patch metric through a set of discontinuous coordinate transformations, valid in regions $r>r_+$, $r_+>r>r_-$ and $r_->r$. For concreteness we will focus on the coordinate transformation valid for $r>r_+$:
\begin{equation}\label{BTZconfcoord}
    \begin{split}
        & x=\sqrt{\frac{r^2-\rp^2}{r^2-\rmm^2}}\cos\Big(\frac{\rp\tau}{L^2}+\frac{|\rmm|\phi}{L}\Big)\exp\Big(\frac{\rp\phi}{L}-\frac{|\rmm|\tau}{L^2}\Big)\\
        & y_E =\sqrt{\frac{r^2-\rp^2}{r^2-\rmm^2}}\sin\Big(\frac{\rp\tau}{L^2}+\frac{|\rmm|\phi}{L}\Big)\exp\Big(\frac{\rp\phi}{L}-\frac{|\rmm|\tau}{L^2}\Big)\\
        & z = \sqrt{\frac{\rp^2-\rmm^2}{r^2-\rmm^2}}\exp\Big(\frac{\rp\phi}{L}-\frac{|\rmm|\tau}{L^2}\Big).
    \end{split}
\end{equation}
Under the coordinate transformation \eqref{BTZconfcoord}, the Euclidean version of the BTZ metric \eqref{BTZmetric} becomes

\begin{equation}\label{PoincarePatch}
    ds^2=\frac{L^2}{z^2}(dx^2+dy_E^2+dz^2).
\end{equation}
We will refer to coordinates like $(x,y_E,z)$ in \eqref{BTZconfcoord} as conformal coordinates, and we will now see that they provide an avenue to analyze the quotient structure of $\mathbb{H}^3/\Gamma$ through a group theoretic lens.

The identification $\phi \sim \phi + 2\pi$ allows for the BTZ black hole to be understood as a quotient of AdS$_{3}$ by a discrete subgroup $\Gamma \sim \mathbb{Z}$ of the  isometry group SL(2,$\mathbb{R})\times$SL(2,$\mathbb{R})$. We can study the group action of the single generator $\gamma\in\Gamma$ by taking $\phi \rightarrow \phi + 2\pi n$ in \eqref{BTZconfcoord}. This will map a point $(x,y_E,z) \in \mathbb{H}^3$ to another point  $(x',y_E',z')$
\begin{equation}
    \gamma^n \cdot (x,y_E,z)=(x^{\prime},y_E^{\prime},z^{\prime})
\end{equation}
through
\begin{equation}\label{BTZmap}
    \begin{split}
        x^{\prime} &= e^{2\pi n \rp / L}\big( x \cos{(2\pi n |\rmm|/L)}-y_E \sin{(2\pi n | \rmm | /L)}\big)\\
        y_E^{\prime} &= e^{2\pi n \rp /L}\big( y_E \cos{(2\pi n |\rmm| /L)} + x\sin{(2\pi n |\rmm|/L)} \big)\\
        z^{\prime} &= e^{2\pi n \rp / L}z.
    \end{split}
\end{equation}
By inspecting \eqref{BTZmap}, it is clear that the group action can be understood as a dilation and a rotation in $\mathbb{R}^2$:
\begin{equation}
    \gamma \begin{pmatrix}
        x \\
        y_E \\
        z \\
    \end{pmatrix}
    =
    \begin{pmatrix}
        e^{2a} & 0 & 0 \\
        0 & e^{2a} & 0 \\
        0 & 0 & e^{2a} \\ 
    \end{pmatrix}
    \begin{pmatrix}
        \cos 2b_E & -\sin 2b_E & 0 \\
        \sin 2b_E & \cos 2b_E & 0 \\
        0 & 0 & 1
    \end{pmatrix}
    \begin{pmatrix}
        x \\
        y_E \\
        z \\
    \end{pmatrix},
\end{equation}
where $a=\pi \rp /L$ and $b_E=\pi |\rmm| /L$.

%We build the subgroup $\Gamma$ by exponentiating the generators of the isometry group, more specifically the killing vector $\partial_{\phi}$. 

We can recast this in the language of \cite{Banados:1992gq}, in which they write the generator $\gamma\in\Gamma$ as the Killing vector  
\begin{equation}
    \partial_{\phi}=\frac{\rp}{L}J_{12} + \frac{|\rmm|}{L}J_{03}.
\end{equation}
Here $J_{12}$ and $J_{03}$ are two of the six generators of the SL(2,$\mathbb{R})\times$SL(2,$\mathbb{R})$ isometry group, and they are presented in terms of $\mathbb{H}^3$ embedding coordinates in Appendix \ref{coordtransfs}. In Poincaré coordinates \eqref{PoincarePatch}, it is evident that these generate dilations $J_{12}=-(x \partial_{x} + y_E \partial_{y_E} + z\partial_{z})$ and rotations $J_{03}=- y_E \partial_{x}+x \partial_{y_E}$. 
In terms of the parameters $a$ and $b_E$, this generator is 
\begin{equation}
   2\pi \partial_{\phi}=2aJ_{12} + 2b_EJ_{03}.
\end{equation}

From the terms $a$ and $b_E$ that parametrize the dilation and rotation, Perry and Williams \cite{perry2003selberg} constructed the Selberg zeta function for the BTZ black hole
\begin{equation}
    Z_{\Gamma}(s)=\prod_{k_{1},k_{2}=0}^{\infty}\left[1-e^{2ib_{E}k_{1}}e^{-2ib_{E}k_{2}}e^{-2a(k_{1}+k_{2}+s)}\right]\;\label{Eulerzeta}.
\end{equation}
In later sections of this article, it will be convenient to consider a Lorentzian version of the zeta function \eqref{Eulerzeta}, which we develop now. To connect to the Lorentzian case we write $y_E\rightarrow iy$, $|\rmm| \rightarrow i \rmm$ and  $b_E\rightarrow-ib$. The group action now consists of a boost in $\bbR^{1,1}$ and a dilation
\begin{equation}
\label{lorentzianquotient}
    \gamma \begin{pmatrix}
        x \\
        y \\
        z \\
    \end{pmatrix}
    =
    \begin{pmatrix}
        e^{2a} & 0 & 0 \\
        0 & e^{2a} & 0 \\
        0 & 0 & e^{2a} \\ 
    \end{pmatrix}
    \begin{pmatrix}
        \cosh 2b & -\sinh 2b & 0 \\
        -\sinh 2b & \cosh 2b & 0 \\
        0 & 0 & 1
    \end{pmatrix}
    \begin{pmatrix}
        x \\
        y \\
        z \\
    \end{pmatrix},
\end{equation}
and the Selberg zeta function is the same as  \eqref{Eulerzeta} but with $i b_{E}=b$
\begin{equation}
    \label{lorentzzeta}
    Z_{\Gamma}(s)=\prod_{k_{1},k_{2}=0}^{\infty}\left[1-e^{2bk_{1}}e^{-2bk_{2}}e^{-2a(k_{1}+k_{2}+s)}\right].
\end{equation}
The exponential part is real in the Lorentzian case because the ``rotation'' part of the quotient is no longer compact.

The zeroes of the zeta function \eqref{lorentzzeta} occur when the exponent is equal to $\pm 2 \pi i k$ for $k \in \mathbb{Z}$. The sign of $k$ differs in the left and right quasinormal modes, hence we have $+$ for the right modes and $-$ for the left modes.
\begin{equation}
    s_{*}=-(k_{1}+k_{2})+ \frac{b}{a}(k_{1}-k_{2}) \pm \frac{i \pi k}{a}.
\end{equation}
Performing a change of basis, the integers $k_{1}$ and $k_{2}$ are written in terms of the radial quantum number $j$ and the thermal integer $n$ in the following way \cite{perry2003selberg}
\begin{equation}
    \label{changeofbasis}
    \begin{split}
        & n \geq 0: \qquad  k_{1}+k_{2}=2j+n \qquad  k_{1}-k_{2}=\mp n,\\
        & n <0: \qquad  k_{1}+k_{2}=2j-n  \qquad k_{1}-k_{2}=\mp n. 
    \end{split}
\end{equation}
For $n \geq 0$ the zeroes are
\begin{equation}\label{ngreatermodes}
    s_* = -(2 j+n)-\frac{b n}{a}-\frac{i \pi  k}{a} \qquad  s_* = -(2 j+n)+\frac{b n}{a}+\frac{i \pi  k}{a}
\end{equation}
and for $n < 0$
\begin{equation}
    s_* = -(2 j-n)-\frac{b n}{a}-\frac{i \pi  k}{a} \qquad  s_* = -(2 j-n)+\frac{b n}{a}+\frac{i \pi  k}{a}.
\end{equation}

\subsection{Warped AdS$_3$ quotients}\label{subsec:2.2warpedbh}

In order to describe warped AdS$_3$ black holes as quotients, we begin by reviewing the global spacetime warped AdS$_3$ itself. Non-warped AdS$_3$ can be expressed as a Hopf fibration of AdS$_2$ like so:
\begin{equation}
  \label{ads3fibred}
  ds^2 = \frac{L^2}{4}(-\cosh^2\sigma d\tau^2 + d\sigma^2+(du + \sinh\sigma d\tau)^2),
\end{equation}
where $u$ is the fibered coordinate and we refer to $(\tau,u,\sigma)$ as global fibered coordinates. Spacelike\footnote{For information on timelike and null analogues, see \cite{Anninos:2008fx}.} warped AdS$_3$ is obtained by warping the fiber length \cite{Bengtsson:2005zj}:
\begin{equation}
  \label{globfibcoord}
  ds^2= \frac{L^2}{\nu^2+3}\left(  -\cosh^2\sigma d\tau^2 + d\sigma^2+ \frac{4\nu^2}{\nu^2+3}(du + \sinh\sigma d\tau)^2 \right).
\end{equation}
When $\nu^2>1$ the spacetime is called \emph{stretched} and for $\nu^2<1$ it is called \emph{squashed}. The Killing vectors for spacelike warped AdS$_3$ are 
\begin{equation}
  \label{wads_kv}
  \begin{split}
    J_2 &= 2\pdd_u \\
    \tilde J_1 &= 2 \sin\tau\tanh\sigma\pdd_\tau - 2\cos\tau\pdd_\sigma+2\sin\tau\sech\sigma\pdd_u \\
    \tilde J_2 &= -2\cos\tau\tanh\sigma \pdd_\tau -2\sin\tau\pdd_\sigma-2\cos\tau\sech\sigma\pdd_u \\
    \tilde J_0 &= 2\pdd_\tau.
  \end{split}
\end{equation}
The tilded vectors form an SL(2,$\mathbb{R}$) algebra, while $J_2$ is a U(1) generator. 

\subsubsection{Spacelike stretched warped AdS$_3$ black holes}

We first consider spacelike stretched warped AdS$_3$ solutions, which are black holes \cite{Anninos:2008fx}. In direct analogy to the BTZ black hole, one can take a quotient of warped AdS$_3$ to find such black hole solutions. The metric for the spacelike warped AdS$_3$ black hole is 
\begin{equation}
  \begin{split}
    \frac{ds^2}{L^2} &= d\tilde t^2 + \frac{d\rho^2}{(\nu^2 + 3)(\rho-\rho_{+})(\rho-\rho_{-})} + \left(2\nu\rho -\sqrt{\rho_{+}\rho_{-}(\nu^2 +3)}\right) d\tilde t d\tilde \phi \\
    &\quad + \frac{\rho}{4}\left(3(\nu^2 - 1)\rho +(\nu^2 + 3)(\rho_{+}+\rho_{-})-4\nu\sqrt{\rho_{+}\rho_{-}(\nu^2 +3)}\right) d\tilde\phi^2, \label{wbhmet}
  \end{split}
\end{equation}
where $\rho \in [0,\infty)$, $\tilde t \in \bbR$, $\tilde \phi \in [0,2\pi)$, and $\tilde \phi \sim \tilde \phi + 2 \pi$.
The coordinate transformation from \eqref{globfibcoord} to \eqref{wbhmet} is given in \eqref{globfibtowbh}. 
For $\nu^2<1$ this spacetime has closed timelike curves for large $r$, but not for $\nu^2>1$. It is important to note that this metric is not asymptotically spacelike warped AdS$_3$ \cite{Anninos:2009zi}.
This can be seen by taking the large $\rho$ limit of the metric \eqref{wbhmet}:
\begin{equation}    
    \label{rotmet}
  \frac{ds^2}{L^2} = d\tilde t ^2 + \frac{d\rho^2}{(\nu^2+3)\rho^2} + 2\nu \rho\, d\tilde t d\tilde \phi + \frac34 (\nu^2 -1)\rho^2 d\tilde\phi^2.
\end{equation}
The angular coordinate $\tilde \phi$ is compact in the asymptotic black hole spacetime \eqref{rotmet}, but it is not in  unquotiented spacelike warped AdS$_3$ (see \cite{Anninos:2009zi} for more details). 

For the spacelike stretched warped AdS$_3$ black hole, the quotient is along the vector: 
\begin{equation}
  \label{quotientvec}
  \pdd_{\tilde\phi} = \pi L (T_L J_2-T_R\tilde J_2),
\end{equation}
where
\begin{equation}
  \label{rhotemps}
  \begin{split}
    T_L &= \frac{\nu^2+3}{8\pi L}\left(\rhp+\rhm -\frac{\sqrt{(\nu^2+3)\rhp\rhm}}{\nu}\right), \qquad
    T_R = \frac{\nu^2+3}{8\pi L}(\rhp-\rhm).
  \end{split}
\end{equation}
For $\nu = 1$, this reduces to the BTZ black hole in a rotated frame $(\tilde t,\rho,\tilde\phi)$.
In order for quantities like $T_L$ and $T_R$ to reduce to their more recognizable BTZ values, it will be convenient to go to the non-rotated frame $(t,r,\phi)$:
\begin{equation}
  \tilde t= \frac{\rp-\rmm}{L^2} t,\quad \tilde \phi = \frac{L \phi - t}{L^2},\quad \rho = \frac{r^2}{\rp-\rmm}, \quad \rho_\pm = \frac{r_\pm^2}{\rp-\rmm}.
\end{equation}
The metric in these coordinates is
\begin{equation}
  \begin{split}
    ds^2 &= \frac{1}{4 L^2 (\rp-\rmm)^2}\bigg(3 r^4(\nu^2 -1) + r^2 \left((\rp^2+\rmm^2)(\nu^2-8\nu+3) - 4\rp\rmm \nu (\sqrt{ \nu^2+3 }-4)\right)\\
    &\qquad + 4(\rp-\rmm)^2\Big(\rp^2+\rmm^2+ \rp \rmm(\sqrt{\nu^2+3}-2)\Big)\bigg) dt^2
    + \frac{4 L^2 r^2 }{(r^2-\rp^2)(r^2-\rmm^2)(\nu^2+3)}dr^2\\
    &\quad -\frac{1}{2 L (\rp-\rmm)^2}\bigg(3r^4(\nu^2-1)+r^2 \left((\rp^2+\rmm^2)(\nu^2-4\nu+3) - 4\rp\rmm \nu \left(\sqrt{\nu^2+3}-2\right)\right)\\
    &\qquad+2\rp\rmm(\rp-\rmm)^2\sqrt{\nu^2+3}\bigg)dtd\phi +\frac{r^2\left(3r^2(\nu^2-1)-4\rp\rmm \nu \sqrt{\nu^2+3}+(\rp^2+\rmm^2)(\nu^2+3)\right)}{4(\rp-\rmm)^2}d\phi^2.
  \end{split}\label{nonrotmet}
\end{equation}
This reduces exactly to the BTZ metric \eqref{BTZmetric} when $\nu=1$. The left and right temperatures in these BTZ-like coordinates are 
\begin{equation}
    \label{rtemps}
        T_L = \frac{\nu^2+3}{8\pi L}\left(\frac{\nu(\rp^2+\rmm^2)-\rp\rmm\sqrt{\nu^2+3}}{\nu(\rp-\rmm)}\right), \qquad
        T_R = \frac{\nu^2+3}{8\pi L}(\rp+\rmm).
\end{equation}
These reduce to the BTZ left and right temperatures when $\nu=1$
\begin{equation}
    T_L = \frac{\rp-\rmm}{2\pi L},\quad T_R = \frac{\rp+\rmm}{2\pi L}.
\end{equation}
\subsubsection{Warped self-dual solutions}
As previously mentioned, spacelike warped AdS$_3$ for $\nu^2<1$ has closed timelike curves (CTCs) for large $r$. Taking a different quotient from \eqref{quotientvec} by identifying along the $J_{2}$ isometry results in no such curves \cite{Anninos:2008fx} and we have the so-called self-dual solutions that are a discrete quotient of warped AdS$_3$ in analogous to the quotients studied in \cite{Coussaert:1994tu}. Self-dual solutions are not black holes in the strictest sense but they can be regarded as such because they have killing horizons in the $(\tilde{t},\rho,\tilde{\phi})$ coordinates and no CTCs. They also have similar thermodynamic behaviours, like an entropy that obeys the Cardy formula, as discussed in \cite{Anninos:2008fx,Chen:2010qm}. 
Self-dual solutions for warped dS$_3$ black holes are studied in \cite{Anninos:2008qb}. 

For the self-dual solutions mentioned in \cite{Anninos:2008fx} the same analysis can be done as for the warped solutions that we will study but with the identification in $\tilde t$ instead of $\tilde \phi$. % (\textcolor{green}{AG:is this true???}) 
We instead adopt a different coordinate system, and look at the self-dual solutions studied in \cite{Chen:2010qm} with identification $\thetat \sim \thetat + 2 \pi \tilde\alpha$. 
%Note that both solutions are locally equivalent to spacelike warped \ads. 
The metric is given by
\begin{equation}
    \label{selfdualoriginal}
    \hspace{-1pt}
    ds^2=\frac{L^2}{\nu^2+3}\bigg(-(\tilde{\rho}-\trhp)(\tilde{\rho}-\trhm)d\taut^2 + \frac{d\tilde{\rho}^2}{(\tilde{\rho}-\trhp)(\tilde{\rho}-\trhm)}+ \frac{4\nu^2}{\nu^2+3}\left(\tilde{\alpha} d\thetat + \frac{2\tilde{\rho} - \trhp - \trhm}{2}d\taut\right)^2 \bigg).
\end{equation}
%where $\tilde{\alpha}$ is a parameter related to the left-temperature of the ``black hole''.
%In the non-rotated frame, the metric becomes
It will be convenient to redefine the radial coordinate to a BTZ-like radial coordinate $\tilde{r}$: 
\begin{equation}\label{singularcoordtransformation}
    \tilde\rho = \frac{\tilde r^2}{\tilde r_+ - \tilde r_-}, \quad \tilde{\rho}_\pm = \frac{\tilde r^2_\pm}{\tilde r_+ - \tilde r_-}. 
\end{equation}
In these coordinates, the line element of the self-dual warped solution is
\begin{equation} \label{selfdualmet}
    \begin{split}
        ds^2=\frac{L^2}{\nu^2+3}&\left(-\frac{ \left(\rt^2-\rmt^2\right) \left(\rt^2-\rpt^2\right)d\taut^2}{(\rmt-\rpt)^2}+\frac{4 \rt^2 d\rt^2}{\left(\rt^2-\rmt^2\right) \left(\rt^2-\rpt^2\right)}
        \right.\\&\qquad\qquad\qquad\qquad\left.
        +\frac{\nu ^2 }{\nu ^2+3}\left(2\tilde{\alpha} d\thetat + \frac{-2 \rt^2+\rmt^2+\rpt^2}{\rmt-\rpt}d\taut \right)^2\right).
    \end{split}
\end{equation}
The left and right temperatures in these coordinates are
\begin{equation} \label{selfdualtemp}
    T_{L}=\frac{\tilde{\alpha}}{2 \pi L}, \qquad T_{R}=\frac{\rpt + \rmt}{4 \pi L}.
\end{equation}
The authors of \cite{Chen:2010qm} comment that in the extremal limit the right temperature $T_R$ vanishes. The reason that this is not apparent in \eqref{selfdualtemp} is that the coordinate transformation \eqref{singularcoordtransformation} is singular in the extremal limit, and thus \eqref{selfdualtemp} holds only in the non-extremal case.

\section{A Selberg zeta function for warped AdS$_3$ black holes}\label{section:SelberglikeWAdS}

Our first step in constructing the Selberg zeta function for warped AdS$_3$ black holes is to identify a suitable set of ``conformal coordinates'' (analogous to equation \eqref{BTZconfcoord} for the BTZ black hole) that will allow us to interpret the identification $\phi\sim\phi+2\pi$ group-theoretically. It is difficult to find a suitable coordinate transformation from the perspective of the metric, since in the warped case we do not have a target Poincar\'e-like metric in mind. That is, for the case of the spacelike warped AdS$_3$ black hole, we seek a coordinate transformation between \eqref{nonrotmet} and a warped version of \eqref{PoincarePatch} (the latter of which we do not have). However, we will now see that we can make progress by borrowing a trick from the hidden conformal symmetry program of \cite{Castro:2010fd} and others, by studying instead the symmetry structure of the scalar wave equation. We outline our general approach below before moving on to the specific examples of spacelike warped AdS$_3$ black holes in Section \ref{section:3.1} and the self-dual solutions in Section \ref{section3.2}.

Consider again the Poincar\'e patch metric of AdS$_3$:
\begin{equation}\label{PoincarePatchW}
    ds^2=\frac{L^2}{z^2}(dw^+dw^-+dz^2).
\end{equation}
The isometry group of \eqref{PoincarePatchW}, SL$(2,\bbR)\times$ SL$(2,\bbR)$, is generated by six Killing vectors:
\begin{equation}
  \label{sl2rgens}
  \begin{split}
    H_1&=i\partial_+, \qquad H_0=i\left(w^+\partial_++\frac{1}{2}z\partial_z\right), \qquad H_{-1}=i((w^+)^2\partial_++w^+z\partial_z-z^2\partial_-),\\
    \Bar{H}_1&=i\partial_-, \qquad \Bar{H}_0=i\left(w^-\partial_-+\frac{1}{2}z\partial_z\right), \qquad \Bar{H}_{-1}=i((w^-)^2\partial_-+w^-z\partial_z-z^2\partial_+).
  \end{split}
\end{equation}
For the case of warped AdS$_3$, the warp parameter $\nu\neq 1$ in \eqref{globfibcoord} breaks the unbarred SL(2,$\mathbb{R}$) symmetry down to a U(1), and thus the Killing vectors of warped  AdS$_3$ are $(H_0, \bar{H}_0,\bar{H}_{\pm 1})$.
Our goal in this section is to construct a locally spacelike warped AdS$_3$ metric in terms of these coordinates, which will be our ``warped Poincar\'e patch.''

Consider for now a massless scalar field $\Phi$ on one of our backgrounds of interest (either the spacelike warped AdS$_3$ black hole or the self-dual solution). In either case, one can show that the solutions of the Klein-Gordon equation are hypergeometric functions (as we will see for the massive scalar case in Section \ref{sec:wave}). Since hypergeometric functions transform in representations of SL(2,$\mathbb{R}$), we conclude that the Klein-Gordon operator is related to the SL(2,$\mathbb{R}$) quadratic Casimir:
\begin{equation}
    \begin{split}
    \mathcal{H}^2&=-\Bar{H}_0^2+\frac{1}{2}(\Bar{H}_1\Bar{H}_{-1}+\Bar{H}_{-1}\Bar{H}_1)\\
                 &=\frac{1}{4}(z^2\partial_z^2-z\partial_z)+z^2\partial_+\partial_-\, .\\
  \end{split}
\end{equation}
In particular, we would like for the Klein-Gordon operator to be proportional to the entire SL(2,$\mathbb{R})\times$U(1) Casimir:
\begin{equation}\label{propcondition}
  (\mc H^2 + \lambda H_0^2)\Phi \propto \covd^2 \Phi
\end{equation}
where the coefficient $\lambda$ of the U(1) generator is yet to be determined.

Our job now is to find a coordinate transformation 
\begin{equation}\label{confcoords}
  \begin{split}
    w^+&%=\sqrt{\frac{r^2-\rp^2}{r^2-\rmm^2}}e^{\alpha\phi+\beta t}
        =\sqrt{\frac{x-\frac{1}{2}}{x+\frac{1}{2}}}e^{\alpha\phi+\beta t},\\
    w^-&%=\sqrt{\frac{r^2-\rp^2}{r^2-\rmm^2}}e^{\gamma\phi+\delta t}
        =\sqrt{\frac{x-\frac{1}{2}}{x+\frac{1}{2}}}e^{\gamma\phi+\delta t},\\
    z&%=\sqrt{\frac{\rp^2-\rmm^2}{r^2-\rmm^2}}e^{1/2((\alpha+\gamma)\phi+(\beta+\delta) t)}
      =\sqrt{\frac{1}{x+\frac{1}{2}}}e^{1/2((\alpha+\gamma)\phi+(\beta+\delta) t)},
  \end{split}
\end{equation}
such that \eqref{propcondition} holds.
In \eqref{confcoords}, $x$ is related to the BTZ-like radial coordinate $r$ in \eqref{nonrotmet} via $x=\frac{r^2-1/2(r_+^2+r_-^2)}{r_+^2-r_-^2}$.
For the field ansatz 
\begin{equation}
    \label{fieldansatz}
    \Phi=R(x)e^{i(k\phi-\omega t)},
\end{equation}
we find that the quadratic Casimir constructed from the conformal coordinates \eqref{confcoords} and generators \eqref{sl2rgens} is
\begin{equation}
  \begin{split}\label{casimireval}
    &(\mathcal{H}^2+\lambda H_0^2)\Phi\\                                       
    &=\left(\partial_x\left(x^2-\frac{1}{4}\right)\partial_x+\frac{(\omega  (\alpha +\gamma )+k (\beta +\delta))^2}{4 \left(x-\frac{1}{2}\right) (\beta  \gamma-\alpha  \delta )^2}-\frac{(\omega  (\alpha -\gamma)+k (\beta -\delta ))^2}{4\left(x+\frac{1}{2}\right) (\beta  \gamma -\alpha \delta )^2}
      + \lambda \frac{(k\delta+\gamma\omega)^2}{(\beta\gamma-\alpha\delta)^2}
      \right)\Phi.
  \end{split}
\end{equation}
Comparing this Casimir to the Klein-Gordon operator will allow us to solve for the conformal coordinate parameters $(\alpha,\beta,\gamma,\delta)$. With those in hand, we will be able to interpret the identification $\phi\rightarrow\phi+2\pi$ group-theoretically, in analogy to \cite{perry2003selberg}, and build a Selberg zeta function for warped black holes. As a nice bonus, we find a warped version of the Poincar\'e patch metric \eqref{PoincarePatchW}, which reduces to \eqref{PoincarePatchW} when $\nu=1$.

Before we move on to specific examples in the next subsections, some comments about \eqref{casimireval} are in order. Hidden conformal symmetry was studied in warped AdS$_3$ black holes in \cite{fareghbal2010hidden} and in self-dual warped solutions in \cite{Li:2010zr}. In both cases, the authors throw away a term in the Klein-Gordon operator, making a ``near-region limit'' or ``soft hair'' argument, in analogy to \cite{Castro:2010fd}. We show below that the terms they throw away are nothing more than the U(1) piece corresponding to our $\lambda$ term in \eqref{casimireval}. We argue that there is no need to take such a limit to discard such terms, as they too are built into the symmetry structure of the Casimir.

\subsection{Warped AdS$_3$ black holes} \label{section:3.1}

The Klein-Gordon operator of the warped AdS$_3$ black hole background \eqref{nonrotmet} is proportional to the quadratic Casimir \eqref{casimireval} with proportionality constant:
\begin{equation}
  \label{propconst}
  \mc H^2 +\lambda H_0^2= \frac{L^2}{\nu^2+3}\covd^2.
\end{equation}
This proportionality constant is determined by comparing the radial derivative pieces.

In $x$ coordinates, the Klein-Gordon operator with the proportionality constant now reads: 
\begin{equation}
    \label{genKGeq}
  \frac{L^2}{3+\nu^2} \covd^2 \Phi = \left(\pdd_x \left(x^2-\frac14\right)\pdd_x + \frac{P}{4\left(x-\frac12\right)}+\frac{Q}{4\left(x+\frac12\right)}+S\right)\Phi
\end{equation}
where
\begin{equation}
  \begin{split}
    P&= \frac{4L^2\left(k \left(2 (\nu -1) r_+^2-2 r_-^2-r_+ r_- \left(\sqrt{\nu ^2+3}-4\right) \right)- L \omega \rp \left(2\nu  r_+-\rmm\sqrt{\nu ^2+3} \right)\right)^2}{(\rp-\rmm)^2(\rp+\rmm)^2(\nu^2+3)^2},\\
    Q&=-\frac{4L^2\left(k\left(-2\rp^2+2(\nu-1)\rmm^2-\rp\rmm(\sqrt{\nu^2+3}-4)\right)-
       L\omega \rmm\left(2\nu\rmm - \rp\sqrt{\nu^2+3}\right)\right)^2}{(\rp-\rmm)^2(\rp+\rmm)^2(\nu^2+3)^2},\\
    S&= \frac{3L^2(\nu^2-1)(k-L\omega)^2}{(\nu^2+3)^2(\rp^2-\rmm^2)^2}.
  \end{split}
\end{equation}
As mentioned previously, in \cite{fareghbal2010hidden} these coefficients are derived in the $(\tilde t,\rho,\tilde\phi)$ coordinates of \eqref{wbhmet}, but the U(1) term, $S$, is thrown away in the spirit of the hidden conformal symmetry program in a certain limit of their eigenvalue $\omega$. 
There is no need to throw away this piece, as it is part of the symmetry group and does not cause any obstruction in building the conformal coordinates. In fact, the U(1) piece will help us choose the proper solution branch, as we will see.

Equating the coefficients of $k$ and $\omega$ in $P$ and $Q$ to those in the $x\pm\frac12$ poles in the Casimir \eqref{casimireval}, one can solve for $(\alpha,\beta,\gamma,\delta)$.
Due to the squares, there are 4 branches of solutions, and we pick the ones which reduce to those found for BTZ when $\nu$ is set to 1 as reported in Appendix \ref{appendix:BTZconfcoord}\footnote{Had we not known what they reduce to, we can use the U(1) term to eliminate 2 of the four branches, and then we pick the positive branch for $\alpha$.}.

They are:
\begin{equation}
    \begin{split}
       \alpha &=\frac{\left(\nu ^2+3\right) \left(\nu (\rp^2+ \rmm^2)-\rp\rmm\sqrt{\nu ^2+3} \right)}{4 L \nu(\rp-\rmm)},\\
       \beta &=\frac{(\rp-\rmm)(\nu^2+3)}{2L^2 \nu}-
       \frac{\left(\nu ^2+3\right) \left(\nu (\rp^2+ \rmm^2)-\rp\rmm\sqrt{\nu ^2+3} \right)}{4 L^2\nu(\rp-\rmm)},\\
       \gamma &=\frac{1}{4L} \left(\nu ^2+3\right)\left(\rp+\rmm\right),\\
       \delta &=-\frac{1}{4L^2} \left(\nu ^2+3\right) \left(\rp+\rmm\right).
    \end{split}
\end{equation}  
Note that $\alpha=2\pi T_L$ and $\gamma=2\pi T_R$ as defined in \eqref{rtemps}.
This also allows us to determine $\lambda$ to be $3(\nu^2-1)/4\nu^2$. 

The conformal coordinates that we have built yield a warped version of the AdS$_3$  Poincar\'e patch. 
In these conformal coordinates, the previously slightly terrible metric \eqref{nonrotmet} turns out to be quite simple:
\begin{equation}\label{warpedlineelement}
    \begin{split}
        ds^2
        &= \frac{4L^2}{(\nu^2+3)^2 z^2}\left((\nu^2+3)dw^+ dw^- + 4\nu^2 dz^2+
        \frac{3(\nu^2-1)w^+}{z^2}(d{w^-}^2+2zdw^-dz)\right).
    \end{split}     
\end{equation}
This reduces to the Poincar\'e patch \eqref{PoincarePatchW} for $\nu=1$, unlike the Poincar\'e coordinates presented in \cite{Anninos:2008fx}. 
The Killing vectors of this metric are:
\begin{equation}
    \begin{split}
        \bar H_1 = i\pdd_-, \qquad \bar H_0 &= i(w^- \pdd_- + \frac12 z \pdd_z), \qquad \bar H_{-1} = i(-z^2 \pdd_+ + {w^-}^2 \pdd_- + w^- z \pdd_z),\\ H_0 &= i(w^+ \pdd_+ + \frac12 z \pdd_z),
    \end{split}
\end{equation}
which are a subset of \eqref{sl2rgens}, as expected. 
The killing vectors corresponding to rotation in the $w^\pm$ plane and dilation are 
\begin{equation}
    \begin{split}
        H_0 - \bar H_0 &= i(w^+ \pdd_+ - w^- \pdd_-), \\
        H_0 + \bar H_0 &= i(w^+ \pdd_+ + w^- \pdd_- + z \pdd_z)
    \end{split}
\end{equation}
respectively.
Using the inverse transformation of \eqref{confcoords}, we can also see that the quotient is generated by the group element 
\begin{equation}
	e^{-i4\pi^2(T_R \bar{H}_0+T_L H_0)}=e^{2\pi\partial_\phi},
\end{equation}
as discussed in \cite{Castro:2010fd}. 
Comparing this with \cite{Anninos:2008fx} and \eqref{quotientvec}, we see 
\begin{equation}
    H_0= \frac{i}{2}J_2 , \quad \Bar{H}_0=-\frac{i}{2}\tilde J_2.
\end{equation}
Now that we have the warped black hole quotient structure and a set of suitable conformal coordinates in hand, we can repeat the analysis of Perry and Williams (outlined in Section \ref{subsec:SelbergBTZ}) to build our Selberg zeta function for warped AdS$_3$ black holes. 

We begin by switching to the coordinates $x=(w^++w^-)/2$ and $y=(w^+-w^-)/2$. 
Under the transformation $\phi\to\phi+2\pi$, we see that the coordinates once again transform in the same way as \eqref{lorentzianquotient}:
\begin{equation}
    \begin{pmatrix}
        x'\\y'\\z'
    \end{pmatrix}=
    \begin{pmatrix}
        e^{2a}&0&0\\0&e^{2a}&0\\0&0&e^{2a}
    \end{pmatrix}
    \begin{pmatrix} 
        \cosh{2b} & -\sinh{2b} & 0 \\
        -\sinh{2b} & \cosh{2b} & 0 \\
        0 & 0 & 1
    \end{pmatrix}
    \begin{pmatrix}
        x\\y\\z
    \end{pmatrix}
\end{equation}
where we have 
\begin{equation}
    \label{warped_ab}
    \begin{split}
        a &= \frac{\pi\rp(\nu^2+3)(2\nu\rp-\rmm\sqrt{\nu^2+3})}{8 L\nu(\rp-\rmm)}\\
        b &= \frac{\pi\rmm(\nu^2+3)(-2\nu\rmm+\rp\sqrt{\nu^2+3})}{8 L\nu(\rp-\rmm)}.
    \end{split}
\end{equation}
The prime geodesic on the quotient spacetime is the curve which remains invariant under the rotation, which is the $x=y=0$ line. The length of this prime geodesic $\ell$ is related to the dilation parameter by $\ell=2a$. 

Finally, the Selberg zeta function for warped quotients that we propose is \eqref{lorentzzeta}, where in the case of warped AdS$_3$ black holes the parameters $a$ and $b$ are \eqref{warped_ab}. We will see in Section \ref{sec5:mappingzeroes} that the zeroes of this zeta function are successfully mapped to the quasinormal modes on the warped AdS$_3$ black hole backgrounds. It should be noted that, in principle, the structure of the Selberg zeta function for warped quotients could have been more complicated (cf. the function $N(p)$ in equation \eqref{selberggeneral}) than in the BTZ case, or indeed it need not have existed at all. We suspect that for more complicated orbifolds $N(p)$ may need to be constructed with greater care. 

\subsection{Warped self-dual solutions}\label{section3.2}

Following the same procedure as in Section \ref{section:3.1} we construct conformal coordinates for the self-dual solutions \eqref{selfdualmet}, using the coordinate transformation
\begin{equation}\label{confcoordssd}
  \begin{split}
    w^+&=\sqrt{\frac{x-\frac{1}{2}}{x+\frac{1}{2}}}e^{\alpha\thetat+\beta \taut},\\
    w^-&=\sqrt{\frac{x-\frac{1}{2}}{x+\frac{1}{2}}}e^{\gamma\thetat+\delta \taut},\\
    z&=\sqrt{\frac{1}{x+\frac{1}{2}}}e^{1/2((\alpha+\gamma)\thetat+(\beta+\delta) \taut)}.
  \end{split}
\end{equation}
The Klein-Gordon operator for this background is proportional to the quadratic Casimir \eqref{casimireval} and has the same proportionality constant as the warped AdS$_{3}$ black hole
\begin{equation}
  \label{propconstsd}
  \mc H^2 +\lambda H_0^2= \frac{L^2}{\nu^2+3}\covd^2.
\end{equation}
Explicitly, in $x$ coordinates and using the ansatz $\Phi=e^{i(k \thetat - \omega \taut)}R(\tilde r)$, we have: 
\begin{equation} \label{genKGeq_sd}
  \frac{L^2}{3+\nu^2} \covd^2 \Phi= \left(\pdd_x \left(x^2-\frac14\right)\pdd_x + \frac{P}{4\left(x-\frac12\right)}+\frac{Q}{4\left(x+\frac12\right)}+S\right)\Phi
\end{equation}
where
\begin{equation}
    \label{Selfdualpoles}
    \begin{split}
         &P=\left(\frac{k}{ \tilde{\alpha}} + \frac{2\omega}{
        (\rpt + \rmt)} \right)^2, \\
        &Q= - \left(\frac{k}{\tilde{\alpha} } - \frac{2 \omega}{ (\rpt + \rmt)} \right)^2,\\
        &S=\frac{3 k^2 \left(\nu ^2-1\right)}{4 \tilde{\alpha}^2 \nu ^2}.
    \end{split}
\end{equation}

Again we equate the coefficients $k$ and $\omega$ in \eqref{Selfdualpoles} to the ones in \eqref{casimireval} and solve for $(\alpha, \beta, \gamma, \delta)$. The U(1) term eliminates two of the four branches and we choose the positive remaining one
\begin{equation}\label{parameters}
    \alpha = \tilde{\alpha}, \qquad \beta=\gamma=0, \qquad \delta=\frac{1}{2}(\rpt + \rmt).
\end{equation}
Note that the coefficient of the U(1) term is the same as for the warped AdS$_3$ black hole: $\lambda=3 \left(\nu ^2-1\right)/4 \nu ^2$. The parameters \eqref{parameters} are related to the temperatures \eqref{selfdualtemp} via $\alpha=2 \pi L T_{L}$ and $\delta=2 \pi L T_{R}$. 
The conformal coordinates \eqref{confcoordssd} with these values of ($\alpha, \beta, \gamma, \delta$) yield the same warped Poincaré patch as \eqref{warpedlineelement}. From the conformal coordinates we calculate the coefficients $a$ and $b$ that appear in the quotient.
Again we switch to $x=(w^++w^-)/2$ and $y=(w^+-w^-)/2$ and under $\thetat \rightarrow \thetat + 2 \pi\tilde\alpha$ the coordinates transform in the same way as \eqref{lorentzianquotient}, and we obtain
\begin{equation} \label{sdabquotient}
    a=\frac{\pi\tilde{\alpha}}{2}, \qquad b=-\frac{\pi\tilde{\alpha}}{2}.
\end{equation}
The quotient is generated by the Killing vector 
\begin{equation}
    \partial_{\thetat}=\pi L T_L J_{2}=\frac{\tilde \alpha}{2}J_2,
\end{equation}
which now can be expressed in terms of the warped Poincar\'e patch generators:
\begin{equation}
    \pdd_{\thetat} = -i \tilde \alpha H_0.
\end{equation}

\section{Scalar fields on WAdS$_3$ black hole backgrounds}\label{sec:wave}
In order to map the zeroes of our Selberg zeta function to the quasinormal modes of the warped AdS$_3$ black holes in Section \ref{sec5:mappingzeroes}, we need to first study several aspects of a massive scalar field propagating on these backgrounds. We begin with the conformal weight $\Delta$ in Section \ref{subsec:4.1}, followed by the thermal (Matsubara) frequencies $\omega_n$ in Section \ref{sec:thermal} and finally quasinormal modes $\omega_*$ in Section \ref{subsec:4.2}. 
\subsection{Conformal weights of the scalar field}\label{subsec:4.1}

Here we compute the conformal weight of highest weight representations of the symmetry algebra of the black hole spacetime. 
We use the relationship between the quadratic Casimir and the Laplacian of the massive scalar \cite{Anninos:2009jt} 
\begin{equation}
    (\mc H^2+\lambda H_0^2)\Phi = \frac{L^2}{\nu^2+3} m^2 \Phi.
\end{equation}
We can express the SL(2,$\bbR)$ Casimir in terms of the SL(2,$\bbR)$ conformal weight 
\begin{equation}
    \mc H^2 \Phi = h(h-1)\Phi,
\end{equation}
and we know from \eqref{genKGeq} that the U(1) generator acting on the scalar field is
\begin{equation}
    \lambda H_0^2 \Phi = S(k,\omega) \Phi.
\end{equation}
Combining the two we have
\begin{equation}
\label{eq:casimereqn}
    h(h-1)= \frac{L^2}{\nu^2+3} m^2 - S.
\end{equation}
Solving for $\Delta=2h$ and choosing the positive branch, we obtain the conformal weight of the scalar field:
\begin{equation}
    \label{masterdeltaeqn}
    \Delta=1+\sqrt{1+4\left(\frac{m^2 L^2}{\nu^2+3}-S\right)}.
\end{equation}
Note that $S$ depends on the choice of coordinates. 
In the $(\tilde t,\rho,\tilde\phi)$ coordinates of \eqref{wbhmet}, we have 
\begin{equation}\label{bhS}
    S = \frac{3L^2(\nu^2 -1)}{\nu^2+3}\omega^2
\end{equation}
which reproduces the result in \cite{Chen:2009hg}. 
In the coordinates of $(t,r,\phi)$ \eqref{nonrotmet}, we have
\begin{equation}
    \label{nonrotS}
    S =  \frac{3L^2(\nu^2 -1)}{\nu^2+3} \left(\frac{k-L \omega}{\rp^2-\rmm^2}\right)^2
\end{equation}
and in the $(\tilde\tau,\tilde r,\tilde\theta)$ coordinates of (\ref{selfdualmet}) for the self-dual solution, we have
\begin{equation}\label{selfdualS}
    S=\frac{3 k^2 \left(\nu ^2-1\right)}{4 \tilde{\alpha}^2 \nu ^2}.
\end{equation}
Equation \eqref{selfdualS} is the same as the one computed for spacelike stretched warped AdS$_3$ with $\tilde \alpha = 1$, which reproduces the result in \cite{Anninos:2008fx}. 

Notice that for large values of $\omega$ or $k$ in \eqref{bhS}, \eqref{nonrotS} and \eqref{selfdualS} the conformal weight becomes complex. The asymptotic behaviour of the solutions ($\sim r^{-\Delta/2}$) implies that these solutions are travelling waves. The fact that these becomes complex indicate that these spacetimes have superradiant behaviour \cite{Anninos:2009jt}. Since $\Re(\Delta)>0$ the outgoing wave solutions are always normalizable \cite{Bardeen:1999px}.
Note that for the negative branch of \eqref{eq:casimereqn} for small enough values of $S$ we have $\Re(\Delta_-) < 0$ which renders the solutions non-normalizable.
We only consider the positive branch since the Selberg zeta function can be expressed a sum over eigenvalues of normalizable functions via the Selberg trace formula\footnote{We thank the referee for bringing this fact to our notice.}.

\subsection{Thermal frequencies} \label{sec:thermal}
A procedure for calculating thermal frequencies in rotating spacetimes was outlined in \cite{Castro:2017mfj}. Due to translation symmetry in both $t$ and $\phi$, we can use the following ansatz for $\Phi$:
\begin{equation}
  \Phi(t,r,\phi)=e^{-i \omega t + i k \phi} f(r). \label{ansatzsc}
\end{equation}
Using this ansatz in the equation of motion, we obtain a second order differential equation in $r$, which we examine around the outer horizon $\rp$:
\begin{equation}
  A(r) f(r) + B(r) (r-\rp) f'(r) + (r-\rp)^2 f''(r)=0.
\end{equation}
For the spacelike warped AdS$_3$ black hole, we have
\begin{equation}
  \begin{split}
    A &= \frac{16 r^2 L^2}{((r+\rp)(r^2-\rmm^2))^2}\bigg(k^2 g_{tt} +  k \omega g_{t\phi} + \omega^2 g_{\phi\phi}\bigg)\\
    B &= \frac{3r^4- \rp^2 \rmm^2 - r^2(\rp^2+\rmm^2)}{r(r^2-\rmm^2)(r+\rp)}.
  \end{split}
\end{equation}

To construct the indicial equation, we must input a power series solution around $\rp$ for $A, B$ and $f$, like so:
\begin{equation}
  \begin{split}
    A= \sum_{n=0}^{\infty} a_n (r-\rp)^n,\quad B= \sum_{n=0}^{\infty} b_n (r-\rp)^n,\quad f=(r-\rp)^\alpha\sum_{n=0}^\infty f_n (r-\rp)^n.
  \end{split}
\end{equation}
We only need the leading terms of $A$ and $B$ for the indicial equation, which after some algebra turn out to be:
\begin{equation}
  \begin{split}
    a_0 &= \left(\frac{L(2\rmm^2 - 2\rp^2(\nu-1)+\rp\rmm(\sqrt{\nu^2+3}-4)) k + L \rp\left(2 \nu \rp -\rmm \sqrt{\nu^2+3}\right) \omega}{(\rp-\rmm)^2(\rp+\rmm)(\nu^2+3)}\right)^{2},\\
    b_0 &= 1.
  \end{split}
\end{equation}
If we set $\alpha=i n/2$ with $n$ an integer, we can solve the indicial equation for $\omega$ in terms of $n$, obtaining the thermal frequencies:
\begin{equation}
  \omega_n = \frac{i n (\rp-\rmm)^2(\rp+\rmm)(\nu^2+3)-2L k (2\rmm^2 + 2\rp^2(1-\nu)+\rp\rmm(\sqrt{\nu^2+3}-4)) }{2L^2\rp\left(2\nu \rp - \rmm \sqrt{\nu^2+3} \right)}.
\end{equation}
If we set $\nu$ to 1, we recover the thermal frequencies for the BTZ black hole \cite{Castro:2017mfj}:
\begin{equation}
  \omega_n = \frac{L k \rmm + i n (\rp^2 - \rmm^2)}{L^2 \rp},
\end{equation}
which are also known as the Matsubara frequencies. In the $(\tilde t,\rho,\tilde\phi)$ coordinates of \eqref{wbhmet}, this result becomes
\begin{equation}
  \omega_n = \frac{-4k-i n(\rhp-\rhm)(\nu^2+3)}{4\rhp \nu -2 \sqrt{\rhp\rhm(\nu^2+3)}}.
\end{equation}
Repeating the analysis in the self-dual solution (\ref{selfdualmet}) we get
\begin{equation}\label{dualthermal}
    \omega_{n}=-\frac{(\rmt +\rpt) (k-i \tilde{\alpha} n)}{2 \tilde{\alpha}}.
\end{equation}

\subsection{Quasinormal modes}\label{subsec:4.2}
To compute quasinormal modes, we must study the equation of motion of the massive scalar field in the black hole background
\begin{equation}
  (\covd^2-m^2) \Phi(t,r,\phi)=0.
\end{equation}
These have been computed in the ``rotated coordinates'' (($\tilde t,\rho,\tilde\phi$) as in \eqref{wbhmet}) before \cite{Chen:2009hg,Ferreira:2013zta}, but here we present the calculation and results in the BTZ-like coordinates (($t,r,\phi$) as in \eqref{nonrotmet}), in which the quasinormal modes reduce to those of the BTZ black hole \cite{Birmingham:2001hc} when the warping parameter $\nu$ is set to 1. 

\subsubsection{Warped AdS$_3$ black holes} \label{section:QNMbtz}
Here we follow the procedure described in \cite{Birmingham:2001hc}. 
The structure of the equation of motion is clear when we make this coordinate change:
\begin{equation}
  \zeta=\frac{r^2-\rp^2}{r^2-\rmm^2},
\end{equation}
and we will use the same ansatz as \eqref{ansatzsc}: $\Phi(t,\zeta,\phi)= e^{-i \omega t + i k \phi} f(\zeta)$. 
In these coordinates we have
\begin{equation}
  \zeta(1-\zeta)f''(\zeta)+(1-\zeta)f'(\zeta)+\left(\frac{A}{\zeta}+B+\frac{C}{1-\zeta}\right)f(\zeta)=0, \label{qnmdiffeq}
\end{equation}
where
\begin{equation}
  \begin{split}
    A&= \frac{L^2 \left(k \left(2 (\nu -1) r_+^2-2 r_-^2-\left(\sqrt{\nu ^2+3}-4\right) r_+ r_-\right)-L r_+ \omega  \left(2\nu  r_+ - r_-\sqrt{\nu ^2+3} \right)\right)^2}{\left(r_--r_+\right)^4 \left(r_-+r_+\right)^2(\nu^2+3)^2},\\
    B&=-A(\rp\leftrightarrow\rmm),\\
    C&=\frac{3 L^2 \left(\nu ^2-1\right) (k-L \omega )^2}{(\nu^2+3)^2\left(r_--r_+\right)^2}-\frac{L^2 m^2}{\nu^2+3}.
  \end{split}
\end{equation}
The solution to \eqref{qnmdiffeq} is
\begin{equation}
  \begin{split}
    f(\zeta)= \zeta^\alpha &(1-\zeta)^\beta \, _2 F_1 (a,b,c;\zeta),\\
    a=\alpha+\beta+i\sqrt{-B},\quad b&=\alpha+\beta-i\sqrt{-B},\quad c=2\alpha+1.
  \end{split}
\end{equation}
The exponents $\alpha$ and $\beta$ describe the behaviour of the scalar field near the outer horizon and at infinity respectively:
\begin{equation}
  \alpha^2 = - A,\quad \beta=\frac12\left(1-\sqrt{1-4C}\right).
\end{equation}
Notice that $\beta$ can be expressed in terms of the conformal weight of the scalar field \eqref{masterdeltaeqn} and \eqref{nonrotS} as 
\begin{equation}\label{betaconformal}
  \beta=1-\frac{\Delta}{2}.
\end{equation}
We require that the solution is ingoing at the horizon\footnote{Modes that are outgoing at the horizon are sometimes called antiquasinormal. These can be treated similarly, but here we just stick to the ingoing modes for simplicity.} and vanishes at infinity, so we set the following boundary conditions\footnote{To see how to derive these conditions look at equation (15) in \cite{Birmingham:2001hc}.}:
\begin{equation}
  c-a=-j,\quad c-b=-j,\quad j\in\bbZ_{\geq 0}.
\end{equation}
Here we can solve these boundary conditions for $\omega$ in terms of $k$ and $\Delta$.
We note that $\Delta$ does depend on $\omega$, but we keep it implicit since the matching of the Selberg zeta function zeroes $s_*$ to the quasinormal modes does not depend upon the intrinsic structure of $\Delta$, as we will see. 
Thus we obtain 
\begin{equation}
  \omega^*_L=  \frac{k}{L} -i\frac{(\Delta+2j)\left(\nu^2+3\right)(\rp-\rmm)}{4 L^2 \nu },
\end{equation}
and 
\begin{equation}
  \omega^*_R=-\frac{k}{L} \left(\frac{2 \left(\rp-\rmm \right)^2}{\nu ( r_+^2 + r_-^2)-r_+ r_-\sqrt{\nu ^2+3}}-1\right)-\frac{i\left(\nu ^2+3\right) \left(\rp+\rmm\right) \left(\rp-\rmm\right)^2 (\Delta +2 j)}{4 L^2 \left(\nu ( r_+^2 + r_-^2)- r_+ r_-\sqrt{\nu ^2+3}\right)}.
\end{equation}
We recover the BTZ quasinormal modes as calculated in \cite{Birmingham:2001hc} upon setting $\nu$ to 1. 

\subsubsection{Warped self-dual solutions}
We compute the quasinormal modes for the self-dual solutions and reproduce previous results \cite{Li:2010zr,Li:2010sv} in $(\taut,\tilde{\rho}, \thetat)$ coordinates. Using the same ansatz: $\Phi(\taut,\zeta,\thetat)= e^{-i \omega \taut + i k \thetat} f(\zeta)$
and change of coordinates
\begin{equation}
    \zeta=\frac{\tilde{r}^2-\rpt^2}{\tilde{r}^2-\rmt^2},
\end{equation}
we reach the same radial equation as before
\eqref{qnmdiffeq}, where 
\begin{equation}
    \begin{split}
        &A=\left(\frac{k}{2 \tilde{\alpha}} + \frac{\omega}{\rpt + \rmt} \right)^2,\\
        &B=-\left(\frac{k}{2 \tilde{\alpha}} - \frac{\omega}{\rpt + \rmt} \right)^2,\\
        &C=\frac{3 k^2 \left(\nu ^2-1\right)}{4 \tilde{\alpha}^2 \nu ^2}-\frac{L^2 m^2}{\nu ^2+3}.
    \end{split}
\end{equation}
Following the procedure done in Section \ref{section:QNMbtz}, with the corresponding $\Delta$ for the self-dual solution, we obtain the quasinormal modes.
The boundary conditions only permit one solution, the right ingoing mode:
\begin{equation}\label{dualqnm}
    \omega^*_R=-\frac{1}{4} i(\rpt+\rmt)(\Delta +2j).
\end{equation}

%The outgoing
%\begin{equation}
%    \omega_{R}=\frac{1}{4} i (\rpt+\rmt)(\Delta+2j)
%\end{equation}

\section{Mapping the Selberg zeroes to quasinormal modes}\label{sec5:mappingzeroes}
Now we take all the results from the previous sections and show that tuning the zeroes of the Lorentzian Selberg zeta function $s_*$ to the conformal weights $\Delta$, we recover the condition that the corresponding quasinormal modes must be tuned to the thermal frequencies, in the spirit of \cite{Keeler:2018lza}. 

For simplicity, we only consider the ingoing (quasinormal) modes, and so we only need to consider the case with $n\geq0$ in the change of bases listed in \eqref{changeofbasis}. It should be noted that the change of basis \eqref{changeofbasis} did not (to our knowledge) necessarily have to be the same in the warped case. Nevertheless, this change of basis turns out to be consistent in mapping the Selberg zeroes to the warped quasinormal modes. Since the basis change \eqref{changeofbasis} is a result from scattering theory on the Euclidean BTZ background \cite{perry2003selberg}, the fact that \eqref{changeofbasis} still works in the warped case probably tells us that warping the fiber length does not significantly change the poles of the scattering matrix.
\subsection{Warped AdS$_3$ black holes}

The procedure outlined in \cite{Keeler:2018lza} was to express $s_*-\Delta$ in terms of the corresponding quasinormal mode and BTZ thermal frequencies. 
Here we will go in the opposite direction and construct $s_*$ from the quasinormal modes, the warped black hole's thermal frequencies and the conformal weight. We will then solve for $a$ and $b$ used in the quotient, and show that they coincide with \eqref{warped_ab} in the warped black hole case and (\ref{sdabquotient}) in the self-dual case.
That is, we write
\begin{equation} \label{obtainingzeroes}
    s_{*L} = \Lambda_L^{-1}(\omega^*_L-\omega_n)+\Delta, \quad s_{*R} = \Lambda_R^{-1}(\omega^*_R-\omega_n)+\Delta, 
\end{equation}
where $\Lambda_L$ and $\Lambda_R$ are proportionality constants, which can be determined by eliminating $\Delta$ on the right hand side of the equations (since the quasinormal modes depend upon $\Delta$ but the zeroes $s_*$ should not explicitly depend upon $\Delta$). 
Plugging everything that we calculated in previous sections into the right hand side of the equations, we get:
\begin{equation}\label{LambdaLeft}
        \frac{\omega^*_L-\omega_n}{\Lambda_L}+\Delta = -2j-\left(\frac{8 i L \nu (\rp-\rmm)}{\rp(\nu^2+3)(2\rp \nu - \rm \sqrt{\nu^2+3})}\right)k-\left(\frac{2\nu(\rp^2-\rmm^2)}{\rp(2\nu\rp - \rmm \sqrt{\nu^2+3})}\right)n,
\end{equation}
and 
\begin{equation}\label{LambdaRight}
        \frac{\omega^*_R-\omega_n}{\Lambda_R}+\Delta = -2j+\left(\frac{8 i L \nu (\rp-\rmm)}{\rp(\nu^2+3)(2\rp \nu - \rm \sqrt{\nu^2+3})}\right)k-\left(\frac{2((\rp^2-\rmm^2)\nu -\rp\rmm\sqrt{\nu^2+3})}{\rp(2\nu\rp - \rmm \sqrt{\nu^2+3})}\right)n,
\end{equation}
where 
\begin{equation}
    \Lambda_L = \frac{i (\rp-\rmm)(\nu^2+3)}{4L^2 \nu}, \quad \Lambda_R = \frac{i (\rp-\rmm)^2(\rp+\rmm)(\nu^2+3)}{4l^2((\rp^2+\rmm^2)\nu-\rp\rmm\sqrt{\nu^2+3})}.
\end{equation}
Equating \eqref{LambdaLeft} and \eqref{LambdaRight} to \eqref{ngreatermodes} (reproduced here for convenience) 
\begin{equation}\label{zeroesagain}
    s_{*L} = -2j - \left(1+\frac{b}{a}\right)n - \frac{i \pi k}{a},\quad
    s_{*R} = -2j - \left(1-\frac{b}{a}\right)n + \frac{i \pi k}{a},
\end{equation}
and solving for $a$ and $b$, we obtain
\begin{equation}
    \begin{split}
        a &= \frac{\pi\rp(\nu^2+3)(2\nu\rp-\rmm\sqrt{\nu^2+3})}{8 L\nu(\rp-\rmm)},\\
        b &= \frac{\pi\rmm(\nu^2+3)(-2\nu\rmm+\rp\sqrt{\nu^2+3})}{8 L\nu(\rp-\rmm)},
    \end{split}
\end{equation}
which reproduces the values obtained from the quotient \eqref{warped_ab}. 
This shows that the proposed Selberg zeta function zeroes for warped AdS$_3$ black holes are successfully mapped to the quasinormal modes.

\subsection{Warped self-dual solutions}
For the self-dual solutions the left quasinormal mode vanishes, and thus we only match one of the zeroes, $s_{*R}$, to the remaining quasinormal mode. Along the lines of \eqref{obtainingzeroes}, we write
\begin{equation}\label{rightzero}
    s_{*R} = \Lambda_R^{-1}(\omega^*_R-\omega_n)+\Delta ,
\end{equation}
and we find that the proportionality constant is
\begin{equation}\label{lambdadual}
    \Lambda_{R}=-\frac{i}{4}(\rpt + \rmt).
\end{equation}
Plugging everything into the right hand side of \eqref{rightzero} (using \eqref{masterdeltaeqn}, \eqref{selfdualS}, \eqref{dualthermal}, \eqref{dualqnm} and \eqref{lambdadual}), we find
\begin{equation}\label{compare}
    s_{*R}=-2 (j+n)-\frac{2 i k}{\tilde{\alpha} }.
\end{equation}
Comparing \eqref{compare} and \eqref{zeroesagain} allows us to solve for $a$ and $b$:
\begin{equation}
    a= \frac{\pi\tilde{\alpha}}{2}, \qquad b=-\frac{\pi\tilde{\alpha}}{2}.
\end{equation}
This matches our previous result from the quotient (\ref{sdabquotient}). Thus, just as in the warped black hole case in the previous section, the zeroes of our proposed Selberg zeta function for the self-dual solutions are successfully mapped to the quasinormal modes.

\section{Discussion}\label{sec:discussion}
 
Using the quotient structure of warped AdS$_3$ black holes \cite{Anninos:2008fx}, we have constructed a Selberg zeta function for this spacetime in the spirit of \cite{perry2003selberg}, providing an example that extends the work of \cite{perry2003selberg} beyond hyperbolic quotients. We have shown that the zeroes of this zeta function are mapped to the scalar quasinormal modes on the warped AdS$_3$ black hole background \cite{Chen:2009hg, Ferreira:2013zta}, in exactly the same way as in the non-warped (BTZ) case \cite{Keeler:2018lza,Aros:2009pg}. We repeat this analysis for the warped self-dual solutions reported in \cite{Anninos:2008fx, Chen:2010qm}, which are obtained from different quotients than the spacelike warped AdS$_3$ black holes and are interesting in their own right. 

Along the way, we develop a method of constructing a warped version of the AdS$_3$ Poincar\'e patch metric, using the symmetry structure of the Klein-Gordon equation and a familiar ansatz for conformal coordinates \cite{Castro:2010fd,Perry:2020ndy}. To the best of our knowledge, our metric \eqref{warpedlineelement} does not appear in the literature. 
This metric could be of interest in the context of the warped AdS/warped CFT correspondence \cite{Hofman:2014loa}. The conformal boundary of the metric \eqref{warpedlineelement} seems to have the correct properties (such as a degenerate metric) required by the geometry on which a boundary warped CFT lives, which is not manifest in other coordinate systems.

Perhaps the most immediate and intriguing direction for future work is to connect the Selberg zeta function to 1-loop determinants on the warped AdS$_3$ black hole background, in a similar fashion as in \cite{Keeler:2018lza}. With this connection in hand, it might be possible to study quantum corrections on the near-horizon extremal Kerr (NHEK) geometry (which is warped AdS$_3$) \cite{Guica:2008mu}, as well as in other contexts where warped AdS$_3$ appears (see the Introduction for more details).

Furthermore, it is extremely tantalyzing to explore whether this Selberg quotient formalism can be related to the recent and very interesting connection between black hole quasinormal modes and quantum Seiberg-Witten curves \cite{Aminov:2020yma,Bianchi:2021mft}. Both this work and \cite{Aminov:2020yma} fix a geometric lens on the problem of computing spectral data, and a connection between them seems both likely and fruitful to persue. Indeed, such a connection was already hinted at in \cite{Tai:2010gd}. 

Hyperbolic quotient spacetimes occur in several other interesting physical systems in which quantum corrections are of interest. One notable example is AdS wormholes (see for example \cite{Krasnov:2000zq}). In fact, it was recently shown in \cite{Chandra:2022fwi} that certain $k$-boundary wormholes were constructed by quotienting AdS by the discrete group $\mathbb{Z}_k$. Since the quotient group that we consider is isomorphic to $\mathbb{Z}$, these wormholes provide a physically interesting and tractable arena to study more complicated quotienting groups. A second notable example in which hyperbolic quotient geometries arise is in the calculation of holographic entanglement R\'enyi entropies \cite{Barrella:2013wja}. In that work, one difficulty was calculating the generators of the quotienting Schottky group $\Gamma$, and most of the time the authors relied on an expansion in small cross-ration. Perhaps now that we have more experience in calculating generators of quotient groups from the bulk perspective, this problem can be revisited with more success. The main challenge in that endeavor would be making a meaningful connection to the boundary CFT data.

There is strong evidence that the Selberg trace formula also has a different and interesting physical interpretation: it is likely the formal like between two well-known methods for computing functional determinants, namely the heat kernel method (see for example \cite{giombi2008one}) and the quasinormal mode method \cite{Denef:2009kn}. This can be seen from how the Selberg trace formula is generally derived from the Selberg zeta function. As explained in \cite{perry2003selberg}, one considers two different representations of the Selberg zeta function: the Euler product \eqref{Eulerzeta} and the Hadamard product
\begin{equation}\label{hadamard}
    Z_\Gamma(s)=e^{Q(s)}\prod_{s_*}(1-s/s_*)e^{s/s_*+1/2\left(s/s_*\right)^2+1/3\left(s/s_*\right)^3},
\end{equation}
where $Z_\Gamma$ is meromorphic and $Q(s)$ is an entire function. The Selberg trace formula is obtained by taking the logarithmic derivative of \eqref{Eulerzeta} and \eqref{hadamard} and equating them.
To extend this to warped AdS$_3$ quotients, the non-trivial interplay between the asymptotics of $\Delta$, $\omega$ and $k$ may be important to interpret through the lens of \cite{Denef:2009kn}, since this is used to determine the UV counterterms of the partition function, like for example in \cite{Arnold:2016dbb}. We leave this interesting and potentially insightful direction for future work.

\section*{Acknowledgements} We are pleased to thank Dionysios Anninos and Lárus Thorlacius for very helpful discussions. 
This research was supported in part by the Icelandic Research Fund under grants 195970-053 
and 228952-051 and by the University of Iceland Research Fund.

\begin{appendices}

\section{Coordinate transformations}
\label{coordtransfs}
Here we list some useful coordinate transformations between some of the metrics used in this work. 
\begin{itemize}
    \item  \textbf{Embedding coordinates.} The hyperbolic 3-space can be written in terms of its embedding 
    \begin{equation}\label{embedmet}
        ds^2=-dU^2+ dV^2 + dX^2 + dY^2
    \end{equation}
    with the constraint
    \begin{equation}
        -U^2+V^2+X^2+Y^2=-L^2.
    \end{equation}
    The following transformation
    \begin{equation}
        x = \frac{Y}{U+X}, \qquad y_E=\frac{V}{U+X}, \qquad z=\frac{L}{U+X},
    \end{equation}
    is used to write the $\mathbb{H}_3$ line element in the Poincaré patch coordinates \eqref{PoincarePatch}. The isometry generators of \eqref{embedmet} are
    \begin{equation}
        J_{AB} = X_B \pdd_A - X_A \pdd_B.
    \end{equation}
    We can write these explicitly in embedding coordinates and Poincar\'e coordinates:
    \begin{equation}
        \begin{split}
            J_{01} = V \pdd_U + U \pdd_V &= - x y_E \pdd_x + \frac12(x^2 - y_E^2 + z^2 +1)\pdd_{y_E} - y_E z \pdd_z, \\
            J_{02} = X \pdd_V - V \pdd_X &= x y_E \pdd_x - \frac12(x^2 - y_E^2 + z^2 -1)\pdd_{y_E} + y_E z \pdd_z, \\ 
            J_{03} = Y \pdd_V - V \pdd_Y &= - y_E \pdd_ x + x \pdd_{y_E}, \\ 
            J_{12} = X \pdd_U + U \pdd_X &= - x \pdd_x - y_E \pdd_{y_E} - z \pdd_z, \\
            J_{13} = Y \pdd_U + U \pdd_Y &= -\frac12(x^2-y_E^2-z^2-1)\pdd_x - xy_E \pdd_{y_E} - xz \pdd_z, \\
            J_{23} = Y \pdd_X - X \pdd_Y &= \frac12(1-x^2+y_E^2+z^2)\pdd_x - xy_E \pdd_{y_E} - xz \pdd_z.\\
        \end{split}
    \end{equation}
    
    \item \textbf{The stretched warped AdS$_3$ black hole.} The coordinate transformation between warped AdS$_3$ in global fibred coordinates \eqref{globfibcoord} and the warped black hole in rotated coordinates \eqref{wbhmet} is:
    \begin{equation}
    \label{globfibtowbh}
    \begin{split}
        \tau&=\tan^{-1}\bigg[\frac{2\sqrt{(\rho-\rho_{+})(\rho-\rho_{-})}}{2\rho-\rhp-\rhm}\sinh\bigg(\frac{1}{4}(\rhp-\rhm)(\nu^{2}+3)\tilde\phi\bigg)\bigg],\\
        \sigma&=\sinh^{-1}\bigg[\frac{2\sqrt{(\rho-\rho_{+})(\rho-\rho_{-})}}{\rhp-\rhm}\cosh\bigg(\frac{1}{4}(\rhp-\rhm)(\nu^{2}+3)\tilde\phi\bigg)\bigg],\\
        u&=\frac{\nu^{2}+3}{4\nu}\bigg[2\tilde t + \Big(\nu(\rhp+\rhm)-\sqrt{\rhp \rhm(\nu^{2}+3)}\Big)\tilde \phi\bigg] \\
        &\qquad +\coth^{-1}\bigg[\frac{2\rho - \rhp - \rhm}{\rhp-\rhm}\coth\bigg(\frac{1}{4}(\rhp-\rhm)(\nu^{2}+3)\tilde \phi\bigg)\bigg].
    \end{split}
\end{equation}
This transformation is valid for $\rho>\rhp$ \cite{Jugeau:2010nq}. For $\rho < \rhp$ we instead have \cite{Anninos:2008fx}:
\begin{equation}
    \begin{split}
        u&=\frac{\nu^{2}+3}{4\nu}\bigg[2\tilde t + \Big(\nu(\rhp+\rhm)-\sqrt{\rhp \rhm(\nu^{2}+3)}\Big)\tilde \phi\bigg]  \\
        &\qquad -\tanh^{-1}\bigg[-\frac{2\rho - \rhp - \rhm}{\rhp-\rhm}\coth\bigg(\frac{1}{4}(\rhp-\rhm)(\nu^{2}+3)\tilde \phi\bigg)\bigg].
    \end{split}
\end{equation}
\item \textbf{The warped self-dual solution.} The coordinate transformation between warped AdS$_3$ in global fibred coordinates \eqref{globfibcoord} and the self-dual solution \eqref{selfdualoriginal}  is \cite{Chen:2010qm}:
\begin{equation}
    \begin{split}
        \tau&=\tan^{-1}\left[\frac{2\sqrt{(\tilde\rho-\tilde\rho_{+})(\tilde\rho-\tilde\rho_{-})}}{2\tilde\rho-\trhp-\trhm}\sinh(\frac{\trhp-\trhm}{2}\tilde\tau)\right],\\
        \sigma&=\sinh^{-1}\left[\frac{2\sqrt{(\tilde\rho-\tilde\rho_{+})(\tilde\rho-\tilde\rho_{-})}}{2\tilde\rho-\trhp-\trhm}\cosh(\frac{\trhp-\trhm}{2}\tilde\tau)\right],\\
        u&= \tilde\alpha \thetat + \tanh^{-1}\left[\frac{2\tilde\rho-\trhp-\trhm}{\trhp-\trhm}\coth(\frac{\trhp-\trhm}{2}\tilde\tau)\right].
    \end{split}
\end{equation}
\end{itemize}
\section{BTZ conformal coordinates}\label{appendix:BTZconfcoord}

In this appendix we illustrate how our method for obtaining conformal coordinates works in the case of the BTZ black hole. We reproduce the Lorentzian version of \eqref{BTZconfcoord}, as expected.

The BTZ metric is given by
\begin{equation}
    ds^2 = -\frac{r^2-\rp^2-\rmm^2}{L^2}dt^2 + \frac{L^2 r^2}{ (r^2-\rp^2)(r^2-\rmm^2)}dr^2+r^2 d\phi^2 - 2\frac{\rp\rmm}{L}dtd\phi.
\end{equation}
For the BTZ black hole, we find the Klein-Gordon operator is proportional to the quadratic Casimir as:
\begin{equation}
  \label{propconstbtz}
  \mc H^2 = \frac{L^2}{4} \covd^2.
\end{equation}
The Klein-Gordon equation with field ansatz \eqref{fieldansatz} takes the form
\begin{equation}
    \label{btzmet}
  \left(\pdd_x \left(x^2-\frac14\right)\pdd_x + \frac{P_{\text{BTZ}}}{4\left(x-\frac12\right)}+\frac{Q_{\text{BTZ}}}{4\left(x+\frac12\right)}\right)\Phi=0,
\end{equation}
where  $x=\frac{r^2-1/2(r_+^2+r_-^2)}{r_+^2-r_-^2}$ and 
\begin{equation}
  \label{BTZKGpoles}
    P_{\text{BTZ}}= \frac{L^2(k \rmm- \omega L \rp)}{(\rp^2-\rmm^2)^2}, \qquad Q_{\text{BTZ}}= -\frac{L^2(k \rp- \omega L \rmm)}{(\rp^2-\rmm^2)^2}.
\end{equation}

When equating the coefficients $k$ and $\omega$ of \eqref{casimireval} to those in \eqref{BTZKGpoles}, one has 4 branches to chose from.
The branch which yields the correct transformation to the Poincar\'e patch \eqref{PoincarePatchW} is the following:
\begin{equation}
  \label{BTZ_confcoord_branch}
  \begin{split}
    \frac{\alpha+\gamma}{\beta\gamma-\alpha\delta}&= \frac{L^2 \rp}{\rp^2-\rmm^2},\qquad
    \frac{\beta+\delta}{\beta\gamma-\alpha\delta}= -\frac{L\rmm}{\rp^2-\rmm^2},\\
    \frac{\alpha-\gamma}{\beta\gamma-\alpha\delta}&= -\frac{L^2\rmm}{\rp^2-\rmm^2},\qquad
    \frac{\beta-\delta}{\beta\gamma-\alpha\delta}= \frac{L\rp}{\rp^2-\rmm^2}.
  \end{split}
\end{equation}
On solving these set of equations, we have
\begin{equation}
  \label{BTZalphas}
  \alpha=L\beta=\frac{\rp-\rmm}{L},\qquad\gamma=-L\delta=\frac{\rp+\rmm}{L}.
\end{equation}
Again, we see that these are related to the left and right temperatures $\alpha=2\pi T_L$ and $\gamma=2\pi T_R$. Plugging these values into \eqref{confcoords}, one obtains the coordinate transformation from the Boyer-Lindquist coordinates of the BTZ black hole to the Poincar\'e patch
\begin{equation}
    \begin{split}
        w^+&=\sqrt{\frac{r^2-\rp^2}{r^2-\rmm^2}}e^{\frac{1}{L^2}(\rp-\rmm)(t+L\phi)},\\
        w^-&=\sqrt{\frac{r^2-\rp^2}{r^2-\rmm^2}}e^{\frac{1}{L^2}(\rp+\rmm)(L\phi-t)},\\
        z&=\sqrt{\frac{\rp^2-\rmm^2}{r^2-\rmm^2}}e^{\frac{1}{L^2}(L \rp \phi- \rmm t)}.
    \end{split}
\end{equation}
These are nothing more than the Lorentzian version of \eqref{BTZconfcoord}.

\end{appendices}

\bibliography{warped}
\bibliographystyle{JHEP}
\end{document}